\documentclass[journal]{IEEEtran}

\usepackage[T1]{fontenc}
\usepackage[utf8]{inputenc}

\usepackage{amsmath,amssymb,amsfonts,amsthm,bm,mathtools}
\usepackage{graphicx}
\usepackage{placeins}
\usepackage{booktabs}
\usepackage{pifont}
\usepackage{enumitem}
\newlist{assumplist}{enumerate}{1}
\setlist[assumplist,1]{
  label=\textbf{(A\arabic*)},
  ref=A\arabic*,
  leftmargin=*,
  align=left,
  itemsep=0.6ex,
  topsep=0.6ex,
  parsep=0pt
}
\usepackage{cite}

\usepackage{algorithm}
\usepackage{algpseudocode} 
\algrenewcommand\algorithmicrequire{\textbf{Input:}}
\algrenewcommand\algorithmicensure{\textbf{Output:}}

\usepackage[hidelinks]{hyperref}

\theoremstyle{plain}
\newtheorem{theorem}{Theorem}
\newtheorem{lemma}{Lemma}
\newtheorem{corollary}{Corollary}
\newtheorem{proposition}{Proposition}

\theoremstyle{remark}
\newtheorem{remark}{Remark}

\theoremstyle{definition}

\newcommand{\E}{\mathbb{E}}
\newcommand{\Var}{\mathrm{Var}}

\newcommand{\cF}{\mathcal{F}}

\newcommand{\pos}[1]{\left[#1\right]_+}
\newcommand{\negp}[1]{\left[#1\right]_-}

\title{Resource-Element Energy Difference for Noncoherent Over-the-Air Federated Learning}
\author{Hao~Chen,~\IEEEmembership{Member,~IEEE,}
        and Zavareh~Bozorgasl,~\IEEEmembership{Member,~IEEE}%
\thanks{H. Chen (haochen@boisestate.edu) and Z. Bozorgasl (zavarehbozorgasl@u.boisestate.edu) are with the Signal, Communication, and Learning Lab (SCALE Lab), Department of Electrical and Computer Engineering, Boise State University, Boise, ID 83712 USA.}%
}

\begin{document}
\maketitle


\begin{abstract}
Over-the-air federated learning (OTA-FL) reduces uplink latency by aggregating client updates directly over the wireless multiple-access channel. Coherent analog aggregation realizes this idea by aligning the phases and amplitudes of simultaneously transmitted waveforms, which typically requires synchronization, instantaneous channel-state information (CSI), phase compensation, and power control. Noncoherent energy detection removes the need for phase-coherent combining, but a single energy measurement is nonnegative and, therefore, cannot represent signed model updates.

This paper introduces resource-element energy difference (REED), a noncoherent physical-layer primitive for continuous signed aggregation. REED maps the positive and negative parts of each real-valued update to transmit energies on paired orthogonal resource elements and estimates the signed sum by subtracting the corresponding received energies. The construction uses slow-timescale calibration of average channel powers, but does not require instantaneous transmitter- or receiver-side CSI or channel inversion. For independent Rayleigh fading, we derive exact first- and second-moment expressions for single-shot REED and for a chip-diverse extension that spreads each coordinate over multiple independently faded paired chips. The resulting variance laws separate fading-induced self-noise, signal--noise interaction, and receiver-noise fluctuation, giving an explicit diversity--resource tradeoff.

We incorporate REED into Federated Averaging (FedAvg) and prove a smooth nonconvex stationarity bound. Under an average-energy-constrained aggregation gain, the REED perturbation is compatible with the standard $O(1/\sqrt{T})$ stationarity scaling. Experiments on MNIST and Fashion-MNIST show that REED with one paired observation per coordinate tracks clean FedAvg in IID regimes, while chip diversity closes most of the gap in the heterogeneous Fashion-MNIST regime. These results support REED as a simple signed-aggregation mechanism for OTA-FL in settings where instantaneous CSI and coherent alignment are impractical or undesirable.
\end{abstract}


\section{Introduction}

Federated learning (FL) enables distributed clients to train a shared model without uploading raw data, but its wireless implementation is often limited by the repeated uplink transmission of high-dimensional model updates. Over-the-air federated learning (OTA-FL) addresses this bottleneck by exploiting the waveform superposition property of the wireless multiple-access channel: when clients transmit simultaneously, the receiver can obtain an aggregate of their local updates without decoding each update separately~\cite{Yang2020FedLearningAirComp,Amiri2020TSP,Zhu2020BAA}.

A large body of OTA-FL work has been developed under a coherent AirComp model, in which analog update symbols are phase- and amplitude-aligned at the receiver to compute a sum or weighted sum~\cite{Yang2020FedLearningAirComp,Amiri2020TSP,Zhu2020BAA,Cao2021JSAC_AirFEELPowerControl,Cao2022JSAC_AirFedAvgPowerControl,Sery2021OTAHeterogeneous,Zhu2024OTAFLOpt}. This model has enabled effective designs based on transceiver scaling, denoising-factor optimization, device selection, and convergence-aware resource allocation. However, in the physical layer, the transmitted waveforms must arrive with a sufficiently compatible timing, carrier frequency, phase, and amplitude so that the complex-valued superposition represents the intended aggregate. Synchronization, CSI acquisition, fading distortion, waveform design, and channel-dependent compensation therefore remain major obstacles to the practical deployment of AirComp and OTA-FL~\cite{Wang2024AirCompSurvey,xiao2024otafl_survey,PerezNeira2025Waveforms}.

This has motivated alternatives that reduce or avoid full coherent alignment. Some retain continuous aggregation while reducing the CSI burden, for example, through adaptive aggregation weights, massive-MIMO effects, receiver-side processing, or digital modulation~\cite{AzimiAbarghouyi2024WAFeL,Wei2022RandomOrth,Razavikia2024qQAM}. Others change the aggregation representation: one-bit and majority-vote OTA schemes compute sign majorities rather than real-valued sums~\cite{bernstein2018signsgd,ZhuDuGunduzHuang2021TWC,sahin2022noncohmv,Miao2024OneBitAirFL}, while classical noncoherent OAC shifts signed values into a nonnegative range before energy detection~\cite{GoldenbaumBocheStanczak2013,GoldenbaumStanczak2014,lee2024tvt_coh_noncoh}. Recent noncoherent learning schemes use more structured representations: NCOTA-DGD employs a nonnegative codebook/simplex construction for channel-induced decentralized consensus~\cite{Michelusi2024}, and NCAirFL combines noncoherent detection with binary dithering of transmitted coordinates and long-term error compensation for centralized AirFL~\cite{NCAirFL2024}. Together, these approaches show that coherent alignment can be relaxed, but typically by changing the weighting, quantization, or representation of the transmitted update.

This paper proposes resource-element energy difference (REED), a simple paired-energy method for noncoherent signed aggregation in OTA-FL. REED represents each scalar update by its positive and negative parts over paired orthogonal resource elements, and the receiver forms the aggregate by subtracting the corresponding received energies. With slow-timescale average channel-power calibration, REED provides a noncoherent estimator for signed model-update aggregation without instantaneous CSIT/CSIR, channel inversion, artificial DC offsets, or sign-only quantization.

The main contributions are summarized as follows.
\begin{itemize}
    \item A paired-energy formulation for signed noncoherent aggregation. REED preserves both sign and magnitude by replacing phase-coherent summation with an energy-difference operation over paired resource elements.

    \item A physical-layer characterization of the REED estimator. Under independent Rayleigh fading, we derive exact first- and second-moment expressions, decompose the variance into fading-induced self-noise, signal--noise interaction, and receiver-noise fluctuation, and extend the analysis to chip-diverse REED with arbitrary deterministic chip-energy weights. We also introduce independent phase dithering to randomize multiuser energy cross terms when multiple coordinates share a coherence block.

    \item A learning-layer analysis and empirical evaluation in FedAvg. We integrate REED into full-participation FedAvg and prove that, for smooth nonconvex objectives, energy-scaled aggregation preserves the standard $O(1/\sqrt{T})$ stationarity scaling. Experiments under IID and heterogeneous data partitions show how chip diversity reduces the performance loss caused by noncoherent aggregation noise.
\end{itemize}

The remainder of the paper is organized as follows. Section~II develops the scalar REED estimator and its chip-diverse extension. Section~III incorporates REED into FedAvg \cite{McMahan2017FedAvg} and establishes convergence guarantees. Section~IV presents the numerical results, and Section~V concludes the paper.

\section{Signed Scalar Aggregation with REED}
\label{sec:reed}

Consider an AirComp uplink with one aggregation server and \(K\) clients transmitting over a shared wireless resource element. For a single scalar coordinate, client \(k\) holds a real value \(u_k\in\mathbb R\), and the server aims to recover the signed sum
\begin{equation}
    s=\sum_{k=1}^{K}u_k .
    \label{eq:scalar_target}
\end{equation}
The corresponding received baseband signal is
\begin{equation}
    y=\sum_{k=1}^{K}h_k x_k+z,
    \label{eq:mac_scalar}
\end{equation}
where \(x_k\) is the transmitted symbol, \(h_k\) is the complex channel coefficient, and \(z\) is receiver noise. In coherent AirComp, channel-dependent transmit or receive processing aligns the effective complex gains so that \(y\), after scaling, represents the signed sum in \eqref{eq:scalar_target}. In the noncoherent setting considered here, the server uses received energies rather than instantaneous phases. This avoids phase-coherent combining, but energy detection alone cannot distinguish positive and negative real inputs. The key design issue is therefore the signed representation used before energy aggregation.

The closest noncoherent learning schemes use more structured representations. NCOTA-DGD embeds the model state through nonnegative coefficients over a codebook/simplex, so that energy superposition estimates channel-weighted sums of those coefficients before reconstructing a decentralized consensus direction~\cite{Michelusi2024}. This construction is well matched to channel-induced consensus, but its energy use is tied to the codebook representation rather than directly to coordinate amplitudes. NCAirFL uses a stochastic one-sided representation: after adding a memory term, each coordinate is multiplied by a shared binary dither and passed through a positive-part map, while the unrepresented component is retained and compensated over future rounds through error feedback~\cite{NCAirFL2024}. Thus, its instantaneous transmission is a dithered nonnegative proxy rather than the signed update itself.

REED uses a direct coordinate-wise alternative. Define
\[
    u_k^+=[u_k]_+,\qquad u_k^-=[u_k]_-,
\]
where \([x]_+=\max\{x,0\}\) and \([x]_-=\max\{-x,0\}\). Then \(u_k=u_k^+-u_k^-\). REED transmits \(u_k^+\) and \(u_k^-\) over paired orthogonal resource elements and estimates the signed sum by subtracting the corresponding received energies. In this way, signed aggregation is reduced to the difference of two nonnegative energy aggregations, preserving real-valued magnitudes without artificial offsets, codebook embeddings, stochastic one-sided transmission, or long-term memory.

\subsection{Basic Paired-Energy Estimator}

For the positive and negative branches, transmitter \(k\) sends
\begin{equation}
    a_{k,+}
    =
    \frac{\sqrt{\eta u_k^+}}{\mu_k}e^{j\phi_{k,+}},
    \qquad
    a_{k,-}
    =
    \frac{\sqrt{\eta u_k^-}}{\mu_k}e^{j\phi_{k,-}},
    \label{eq:reed_symbols}
\end{equation}
where \(\eta>0\) is an aggregation gain, \(\mu_k^2=\mathbb E|h_k|^2\) is the average channel power, and the dithering phases satisfy
\[
    \phi_{k,\pm}\sim\operatorname{Unif}[0,2\pi),
\]
independently across clients and branches. The receiver observes
\begin{equation}
    y_\pm
    =
    \sum_{k=1}^{K}h_{k,\pm}a_{k,\pm}+z_\pm,
    \qquad
    z_\pm\sim\mathcal{CN}(0,\sigma_z^2),
    \label{eq:reed_received_pair}
\end{equation}
and forms
\begin{equation}
    \hat s
    =
    \frac{|y_+|^2-|y_-|^2}{\eta}.
    \label{eq:reed_estimator}
\end{equation}
The normalization in \eqref{eq:reed_symbols} uses average channel powers only; no instantaneous fading coefficient is estimated or inverted.

\subsection{Moment Characterization}

The paired-energy estimator is unbiased under average channel-power calibration. Define
\[
    S_+=\sum_{k=1}^{K}u_k^+,
    \qquad
    S_-=\sum_{k=1}^{K}u_k^- .
\]

\begin{proposition}[Unbiasedness and variance of the paired-energy estimator]
\label{prop:reed_moments}
Assume \(h_{k,\pm}\sim\mathcal{CN}(0,\mu_k^2)\) independently across transmitters and branches, and assume \(z_+\) and \(z_-\) are independent with variance \(\sigma_z^2\). Then
\begin{equation}
    \mathbb E[\hat s]=s,
    \label{eq:reed_unbiased}
\end{equation}
and
\begin{equation}
    \operatorname{Var}(\hat s)
    =
    S_+^2+S_-^2
    +
    \frac{2\sigma_z^2}{\eta}\sum_{k=1}^{K}|u_k|
    +
    \frac{2\sigma_z^4}{\eta^2}.
    \label{eq:reed_variance}
\end{equation}
\end{proposition}
The proof is given in Appendix~\ref{app:reed_scalar_proofs}.

The variance in \eqref{eq:reed_variance} has three components. The terms \(S_+^2+S_-^2\) are fading-induced self-noise and remain even when receiver noise is negligible. The term proportional to \(\sigma_z^2\) is due to signal--noise interaction, and the term proportional to \(\sigma_z^4\) is receiver-noise energy fluctuation. Increasing \(\eta\) suppresses the receiver-noise terms but not the fading self-noise.

\begin{remark}[Beyond Rayleigh fading]
\label{rem:beyond_rayleigh}
The Rayleigh assumption is used to obtain the compact variance expression in
\eqref{eq:reed_variance}. The same calculation extends to any independent
zero-mean proper fading model with finite fourth moments. Let
\[
    \kappa_k
    \triangleq
    \frac{\mathbb E|h_k|^4}{\left(\mathbb E|h_k|^2\right)^2},
\]
and assume, for notational simplicity, that the positive and negative branches
have the same fourth-moment ratio. Then the paired-energy estimator remains
unbiased, and its variance is
\[
\begin{aligned}
    \operatorname{Var}(\hat s)
    =
    &\,S_+^2+S_-^2
    +
    \sum_{k=1}^{K}(\kappa_k-2)
    \left((u_k^+)^2+(u_k^-)^2\right)  \\
    &+
    \frac{2\sigma_z^2}{\eta}\sum_{k=1}^{K}|u_k|
    +
    \frac{2\sigma_z^4}{\eta^2}.
\end{aligned}
\]
Rayleigh fading gives \(\kappa_k=2\), so the additional fourth-moment term vanishes and \eqref{eq:reed_variance} is recovered. Consequently, the FedAvg convergence analysis in Section~\ref{sec:convergence} applies to these fading models after replacing the Rayleigh value of \(\sigma_{\mathrm{air}}^2\) by the corresponding fourth-moment expression or bound.
\end{remark}

\subsection{Chip-Diverse REED}
\label{sec:chip_reed}

Fading self-noise can be reduced by spreading each scalar over multiple independently faded resource-element pairs. Let \(M\) be the number of chip pairs, let \(c_m\ge 0\) be deterministic chip-energy weights, and define
\[
    C_M=\sum_{m=1}^{M}c_m>0 .
\]
On chip \(m\), transmitter \(k\) sends
\begin{equation}
    a_{k,m,+}
    =
    \frac{\sqrt{\eta c_m u_k^+}}{\mu_k}e^{j\phi_{k,m,+}},
    \qquad
    a_{k,m,-}
    =
    \frac{\sqrt{\eta c_m u_k^-}}{\mu_k}e^{j\phi_{k,m,-}} .
    \label{eq:chip_reed_symbols}
\end{equation}
The received signals are
\begin{equation}
    y_{m,\pm}
    =
    \sum_{k=1}^{K}h_{k,m,\pm}a_{k,m,\pm}+z_{m,\pm},
    \label{eq:chip_reed_received}
\end{equation}
and the receiver estimates
\begin{equation}
    \hat s_M
    =
    \frac{1}{\eta C_M}
    \sum_{m=1}^{M}
    \left(|y_{m,+}|^2-|y_{m,-}|^2\right).
    \label{eq:chip_reed_estimator}
\end{equation}

\begin{proposition}[Unbiasedness and variance of the chip-diverse estimator]
\label{prop:chip_reed_moments}
Assume independent Rayleigh fading, receiver noise, and dithering phases across transmitters, branches, and chips. Then
\begin{equation}
    \mathbb E[\hat s_M]=s,
    \label{eq:chip_reed_unbiased}
\end{equation}
and
\begin{equation}
\begin{aligned}
    \operatorname{Var}(\hat s_M)
    =
    &\frac{\sum_{m=1}^{M}c_m^2}{C_M^2}
    \left(S_+^2+S_-^2\right)
    +
    \frac{2\sigma_z^2}{\eta C_M}
    \sum_{k=1}^{K}|u_k|  \\
    &+
    \frac{2M\sigma_z^4}{\eta^2 C_M^2}.
\end{aligned}
    \label{eq:chip_reed_variance}
\end{equation}
\end{proposition}
The proof is given in Appendix~\ref{app:reed_scalar_proofs}.

Equation~\eqref{eq:chip_reed_variance} separates diversity from additional transmit energy. If the total energy per scalar is fixed and split evenly across chips, then \(c_m=1/M\) and \(C_M=1\), giving
\begin{equation}
    \operatorname{Var}(\hat s_M)
    =
    \frac{S_+^2+S_-^2}{M}
    +
    \frac{2\sigma_z^2}{\eta}\sum_{k=1}^{K}|u_k|
    +
    \frac{2M\sigma_z^4}{\eta^2}.
    \label{eq:chip_fixed_total}
\end{equation}
The fading self-noise decreases with \(M\), while the pure receiver-noise term increases because more noisy energy measurements are accumulated. If each chip uses the same energy as the single-shot transmission, then, \(c_m=1\) and \(C_M=M\), yielding
\begin{equation}
    \operatorname{Var}(\hat s_M)
    =
    \frac{1}{M}
    \left(
    S_+^2+S_-^2
    +
    \frac{2\sigma_z^2}{\eta}\sum_{k=1}^{K}|u_k|
    +
    \frac{2\sigma_z^4}{\eta^2}
    \right).
    \label{eq:chip_fixed_per_chip}
\end{equation}
This gives a full \(1/M\) variance reduction, at the cost of \(M\)-fold transmit energy and \(2M\) resource elements per scalar.

\begin{remark}[Time, frequency, and spatial diversity]
\label{rem:simo_mimo_diversity}
The chip index \(m\) can represent independent time, frequency, or spatial observations. For example, with \(R\) receive antennas,
\[
    y_{r,\pm}
    =
    \sum_{k=1}^{K}h_{k,r,\pm}a_{k,\pm}+z_{r,\pm},
    \qquad r=1,\ldots,R,
\]
and the receive-diverse estimator
\[
    \hat s_R
    =
    \frac{1}{\eta R}
    \sum_{r=1}^{R}
    \left(|y_{r,+}|^2-|y_{r,-}|^2\right)
\]
has, under independent Rayleigh fading across receive antennas,
\[
    \operatorname{Var}(\hat s_R)
    =
    \frac{1}{R}
    \left(
    S_+^2+S_-^2
    +
    \frac{2\sigma_z^2}{\eta}\sum_{k=1}^{K}|u_k|
    +
    \frac{2\sigma_z^4}{\eta^2}
    \right).
\]
Thus, SIMO reception provides the same noncoherent diversity gain as fixed-per-chip repetition, but through receiver hardware rather than additional time--frequency resources or repeated transmit energy. Spatial correlation or unequal branch powers reduce the effective diversity order and change the constants in \eqref{eq:chip_reed_variance}.
\end{remark}

\begin{remark}[Phase dithering across coordinates]
\label{rem:phase_dithering}
The scalar moment laws assume independent effective energy observations across coordinates. When several coordinates share a coherence block, common fading coefficients can create structured multiuser cross terms in the received energies. By drawing the dithering phases independently across resource elements, REED randomizes these cross terms and reduces the structured coupling among scalar estimates that share the same coherence block.
\end{remark}

\section{REED-Enabled FedAvg and Convergence}
\label{sec:convergence}

We consider FedAvg with \(K\) clients. Client \(k\) has local empirical risk
\begin{equation}
    f_k(w)
    =
    \frac{1}{|\mathcal D_k|}
    \sum_{\xi\in\mathcal D_k}\ell(w;\xi),
\end{equation}
and the global objective is
\begin{equation}
    \min_{w\in\mathbb R^d}
    F(w)
    \triangleq
    \frac{1}{K}\sum_{k=1}^{K} f_k(w).
    \label{eq:fedavg_objective}
\end{equation}
At round \(t\), the server broadcasts \(w^t\). Client \(k\) initializes \(w_{k,0}^t=w^t\) and performs \(Q\) local stochastic-gradient steps as follows
\begin{equation}
    w_{k,q+1}^t
    =
    w_{k,q}^t-\beta g_{k,q}^t,
    \qquad q=0,\ldots,Q-1,
    \label{eq:local_sgd}
\end{equation}
where \(\beta>0\) is the local stepsize and
\[
    \mathbb E[g_{k,q}^t\mid w_{k,q}^t]
    =
    \nabla f_k(w_{k,q}^t).
\]
The local model increment is
\begin{equation}
    \Delta_k^t
    \triangleq
    w_{k,Q}^t-w^t
    =
    -\beta\sum_{q=0}^{Q-1}g_{k,q}^t.
    \label{eq:local_increment}
\end{equation}
With ideal communication, the server would apply
\begin{equation}
    \bar\Delta^t
    \triangleq
    \frac{1}{K}\sum_{k=1}^{K}\Delta_k^t .
    \label{eq:ideal_fedavg_increment}
\end{equation}

REED is applied coordinate-wise to estimate \(\bar\Delta^t\). For coordinate \(j\), client \(k\) uses the scalar input
\begin{equation}
    u_{k,j}^t
    =
    \frac{1}{K}[\Delta_k^t]_j,
    \label{eq:reed_fedavg_scalar_input}
\end{equation}
so that \(\sum_k u_{k,j}^t=[\bar\Delta^t]_j\). Let \(\widehat{\Delta}^t\) denote the vector obtained by applying REED to all coordinates. The server update is
\begin{equation}
    w^{t+1}
    =
    w^t+\widehat{\Delta}^t
    =
    w^t+\bar\Delta^t+\varepsilon^t,
    \label{eq:reed_fedavg_update}
\end{equation}
where
\[
    \varepsilon^t
    \triangleq
    \widehat{\Delta}^t-\bar\Delta^t
\]
is the REED aggregation error.

\begin{figure*}[!t]
\centering
\includegraphics[width=0.98\textwidth]{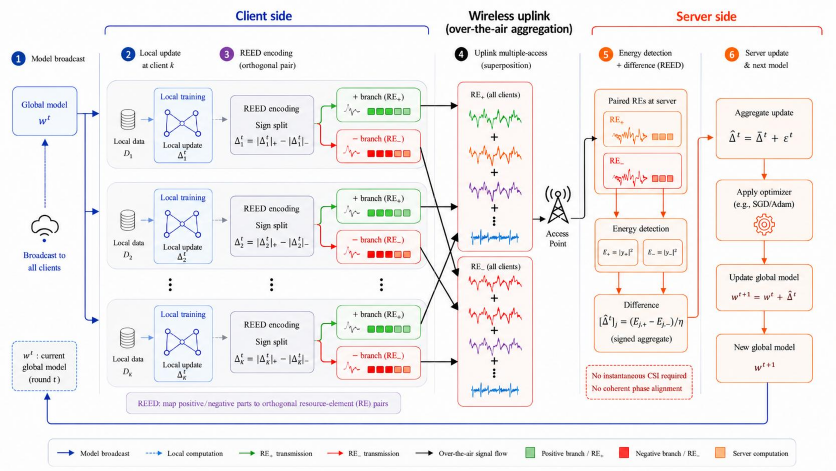}
\vspace{-2mm}
\caption{REED-enabled OTA-FL workflow. Clients compute local FedAvg increments, transmit positive and negative coordinate parts over paired resource elements, and the server forms the signed aggregate through noncoherent energy differencing.}
\label{fig:reed_system_overview}
\end{figure*}

Figure~\ref{fig:reed_system_overview} summarizes one communication round of REED-enabled OTA-FL. The server first broadcasts the current global model \(w^t\). Each client performs local training to form a model increment \(\Delta_k^t\), applies the REED positive--negative split coordinate-wise, and transmits the two branches over paired orthogonal resource elements. The wireless uplink superposes the client waveforms noncoherently. At the server, the energies of each positive--negative resource pair are measured and subtracted, producing an estimate \(\widehat{\Delta}^t\) of the ideal FedAvg increment \(\bar\Delta^t\). The server then updates the global model as \(w^{t+1}=w^t+\widehat{\Delta}^t\). 

We next establish convergence guarantees for this update model under standard assumptions on smoothness, stochastic-gradient noise, and bounded local updates.


\subsection{Assumptions}

Let \(\mathcal F_t\) denote the sigma-field generated by the global model and all randomness up to the beginning of round \(t\). The convergence analysis uses the following assumptions.

\begin{assumplist}
\item \textbf{Smoothness and lower bound.}
Each \(f_k\) is \(L\)-smooth; hence, \(F\) is \(L\)-smooth. Also, \(F(w)\ge F_\star\) for all \(w\).

\item \textbf{Unbiased stochastic gradients.}
For all \(k,q,t\),
\[
    \mathbb E[g_{k,q}^t\mid w_{k,q}^t]
    =
    \nabla f_k(w_{k,q}^t).
\]

\item \textbf{Bounded stochastic gradients.}
There exists \(G<\infty\) such that
\[
    \|g_{k,q}^t\|\le G
\]
almost surely for all \(k,q,t\). Consequently,
\[
    \|\Delta_k^t\|\le \beta QG .
\]

\item \textbf{Gradient-noise variance.}
There exists \(\sigma_g^2<\infty\) such that
\[
    \mathbb E\!\left[
    \|g_{k,q}^t-\nabla f_k(w_{k,q}^t)\|^2
    \mid w_{k,q}^t
    \right]
    \le \sigma_g^2 .
\]
Gradient noises from different clients are conditionally uncorrelated given the local iterates.

\item \textbf{REED communication randomness.}
Coordinate-wise REED uses the channel and dithering assumptions of Section~\ref{sec:reed}. Conditioned on the transmitted increments, the aggregation error is unbiased and has a bounded second moment determined by the chosen REED configuration.
\end{assumplist}

\subsection{Vector Aggregation Error}

The scalar moment laws in Section~\ref{sec:reed} induce a common vector-level perturbation model. The convergence proof only uses unbiasedness and a second-moment bound; the specific REED configuration determines the value of this bound.

\begin{lemma}[REED aggregation error]
\label{lem:reed_vector}
Assume coordinate-wise REED aggregation with equal FedAvg weights as in \eqref{eq:reed_fedavg_scalar_input}. Conditioned on the local increments,
\begin{equation}
    \mathbb E[
    \varepsilon^t
    \mid
    \mathcal F_t,\{\Delta_k^t\}_{k=1}^{K}
    ]
    =
    0,
    \qquad
    \mathbb E[\varepsilon^t\mid\mathcal F_t]=0 .
    \label{eq:reed_error_unbiased_vector}
\end{equation}
Moreover, if the local increments satisfy Assumption~(A3), then
\begin{equation}
    \mathbb E\|\varepsilon^t\|^2
    \le
    \sigma_{\mathrm{air}}^2,
    \label{eq:reed_error_second_moment_generic}
\end{equation}
where a valid choice for REED with one paired observation per coordinate is
\begin{equation}
    \sigma_{\mathrm{air}}^2
    =
    (\beta QG)^2
    +
    \frac{2\sigma_z^2\sqrt d}{\eta}(\beta QG)
    +
    \frac{2d\sigma_z^4}{\eta^2},
    \label{eq:sigma_air}
\end{equation}
and, for chip-diverse REED with weights \(\{c_m\}_{m=1}^{M}\) and \(C_M=\sum_m c_m\), a valid choice is
\begin{equation}
\begin{aligned}
    \sigma_{\mathrm{air}}^2
    =\;&
    \frac{\sum_{m=1}^{M}c_m^2}{C_M^2}(\beta QG)^2
    +
    \frac{2\sigma_z^2\sqrt d}{\eta C_M}(\beta QG) \\
    &+
    \frac{2dM\sigma_z^4}{\eta^2 C_M^2}.
\end{aligned}
\label{eq:chip_sigma_air}
\end{equation}
\end{lemma}
The proof is given in Appendix~\ref{app:convergence_proof}.

\subsection{Aggregation Gain Under an Average-Energy Constraint}

The aggregation gain \(\eta\) controls the receiver-noise terms in \eqref{eq:sigma_air} and \eqref{eq:chip_sigma_air}. Under bounded updates, it can be chosen from a per-round average-energy constraint.

\begin{lemma}[Feasible energy-scaled gain]
\label{lem:eta_beta_scaling}
Assume the bounded-gradient condition in Assumption~(A3). For chip-diverse REED with deterministic weights \(\{c_m\}_{m=1}^{M}\) and \(C_M=\sum_m c_m\), suppose client \(k\) has the per-round average energy constraint
\begin{equation}
    \frac{1}{d}
    \sum_{j=1}^{d}\sum_{m=1}^{M}
    \left(
    |a_{k,j,m,+}^t|^2+
    |a_{k,j,m,-}^t|^2
    \right)
    \le E_k .
    \label{eq:avg_energy_constraint}
\end{equation}
Then, the deterministic choice
\begin{equation}
    \eta
    =
    \min_{k\in\{1,\ldots,K\}}
    \frac{E_k K\sqrt d\,\mu_k^2}{C_M\beta QG}
    \label{eq:eta_choice}
\end{equation}
satisfies all client energy constraints. Hence, \(\eta=\Theta(1/\beta)\) for fixed system parameters and fixed chip weights, and
\begin{equation}
    \sigma_{\mathrm{air}}^2=O(\beta^2)
    \label{eq:sigma_air_beta_scaling}
\end{equation}
for both \eqref{eq:sigma_air} and \eqref{eq:chip_sigma_air}.
\end{lemma}
The proof is given in Appendix~\ref{app:convergence_proof}.

\subsection{Nonconvex Convergence}

\begin{theorem}[FedAvg with REED]
\label{thm:conv}
Under Assumptions (A1)--(A5), suppose the REED aggregation error satisfies Lemma~\ref{lem:reed_vector} with second-moment bound \(\sigma_{\mathrm{air}}^2\). If
\[
    \beta\le \frac{1}{8LQ},
\]
then, for any \(T\ge 1\),
\begin{equation}
\begin{aligned}
    \frac{1}{T}
    \sum_{t=0}^{T-1}
    \mathbb E\|\nabla F(w^t)\|^2
    \le\;&
    \frac{4(F(w^0)-F_\star)}{\beta QT}
    +
    2L^2\beta^2Q^2G^2  \\
    &+
    8L^3\beta^3Q^3G^2
    +
    \frac{4L\beta Q}{K}\sigma_g^2
    +
    \frac{2L}{\beta Q}\sigma_{\mathrm{air}}^2 .
\end{aligned}
\label{eq:conv_bound}
\end{equation}
\end{theorem}
The proof is given in Appendix~\ref{app:convergence_proof}.

The bound in \eqref{eq:conv_bound} has the same structure as a standard smooth nonconvex FedAvg bound, with an additional term
\[
    \frac{2L}{\beta Q}\sigma_{\mathrm{air}}^2
\]
capturing the REED aggregation error. Under the energy-scaled gain in Lemma~\ref{lem:eta_beta_scaling}, \(\sigma_{\mathrm{air}}^2=O(\beta^2)\), so this additional term scales as \(O(\beta)\). Thus, with the standard diminishing stepsize \(\beta=\Theta(T^{-1/2})\), the following nonconvex stationarity guarantee follows.

\begin{corollary}[Nonconvex stationarity guarantee]
\label{cor:stationarity_scaling}
Under the conditions of Theorem~\ref{thm:conv} and Lemma~\ref{lem:eta_beta_scaling}, choosing
\[
    \beta=\Theta(T^{-1/2})
\]
yields
\begin{equation}
    \frac{1}{T}
    \sum_{t=0}^{T-1}
    \mathbb E\|\nabla F(w^t)\|^2
    =
    O(T^{-1/2}),
    \label{eq:stationarity_rate}
\end{equation}
up to the dependence on \(Q\), \(K\), the stochastic-gradient variance, and the constants in the REED second-moment bound.
\end{corollary}

\section{Experimental Results}
\label{sec:experiments}

We evaluate REED as an aggregation method for FedAvg over fading channels. The experiments first compare REED with one paired observation per coordinate (\(M=1\)) against clean FedAvg and coherent CSIT aggregation, and then examine the effect of chip diversity (\(M\in\{2,4\}\) ). All wireless results use an effective receive SNR of \(-10\)~dB, representing a low-SNR operating regime in which noncoherent energy fluctuations are expected to be visible.

\subsection{Experimental Setup}

We use full-participation FedAvg with \(K=10\) clients. Each communication round consists of \(Q=10\) local minibatch SGD steps per client, with minibatch size 64. The local learning rate is
\[
    \beta_t=\frac{0.05}{\sqrt{1+t}},
\]
where \(t\) is the communication-round index. 

Experiments are conducted on MNIST~\cite{lecun1998gradient} and Fashion-MNIST~\cite{xiao2017fashion} using a compact CNN with two Conv--ReLU--MaxPool blocks followed by two fully connected layers and a 10-class output layer.

Experiments are conducted on MNIST and Fashion-MNIST using a compact CNN with two Conv--ReLU--MaxPool blocks followed by two fully connected layers and a 10-class output layer. For both datasets, we use the standard \(60{,}000/10{,}000\) train/test split: the full training set is partitioned across clients, and the standard test set is used for evaluation.

We consider IID and heterogeneous client partitions. In the IID setting, the training indices are randomly shuffled and split approximately evenly across clients. In the heterogeneous setting, we use a label-aware Dirichlet split: for each class, client proportions are drawn from a symmetric Dirichlet distribution with concentration parameter \(\alpha=0.3\), and samples of that class are allocated accordingly. All experiments were repeated over 10 independent trials, with random seeds matched across aggregation schemes within each trial. The reported curves and table entries show the average over these trials.

The baselines are as follows. Clean FedAvg uses ideal noiseless aggregation. Coherent CSIT aggregation is a wireless analog reference with channel compensation for coherent summation. REED with \(M=1\) uses one positive--negative resource-element pair per scalar coordinate in each communication round. Chip-diverse REED uses \(M\in\{2,4\}\) independently faded paired chips, as in Section~\ref{sec:chip_reed}. Here \(M\) denotes the number of paired energy observations per scalar coordinate per communication round.

For REED, the uplink follows the Rayleigh-fading model in Section~\ref{sec:reed}, with additive white Gaussian noise and independent phase dithers across users, signs, coordinates, and chips. The paired resource elements use long-term average channel-power normalization. The reported chip-diverse results use the same per-chip receive-SNR convention as the \(M=1\) setting; they therefore represent diversity-and-resource operating points rather than fixed-total-energy ablations.

\noindent\textbf{Code availability:}
The source code used to reproduce the experiments is available at
\url{https://github.com/zavareh1/REED}.

\begin{figure}[!t]
\centering
\includegraphics[width=\columnwidth]{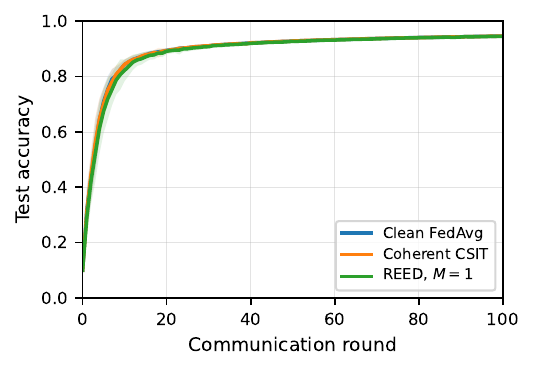}\\[-1mm]
{\footnotesize (a) IID partition.}\\[1mm]
\includegraphics[width=\columnwidth]{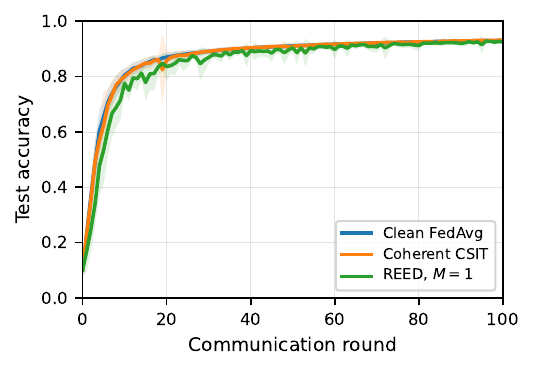}\\[-1mm]
{\footnotesize (b) Dirichlet \(\alpha=0.3\) partition.}
\caption{REED with \(M=1\) on MNIST at \(-10\)~dB effective receive SNR.}
\label{fig:single_shot_mnist}
\end{figure}

\begin{figure}[!t]
\centering
\includegraphics[width=\columnwidth]{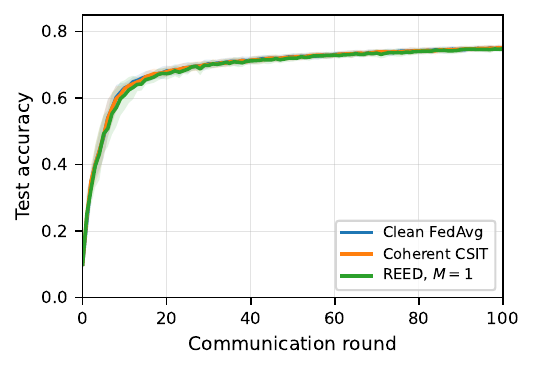}\\[-1mm]
{\footnotesize (a) IID partition.}\\[1mm]
\includegraphics[width=\columnwidth]{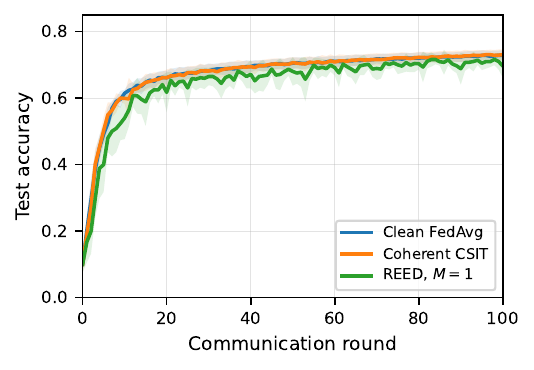}\\[-1mm]
{\footnotesize (b) Dirichlet \(\alpha=0.3\) partition.}
\caption{REED with \(M=1\) on Fashion-MNIST at \(-10\)~dB effective receive SNR.}
\label{fig:single_shot_fashion}
\end{figure}

\begin{figure}[!t]
\centering
\includegraphics[width=\columnwidth]{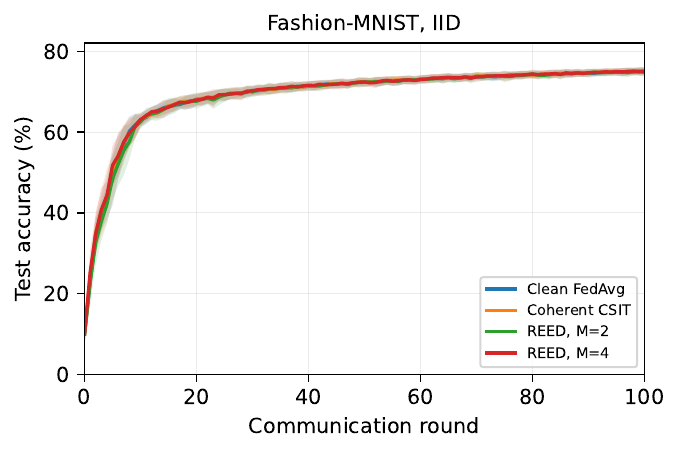}\\[-1mm]
{\footnotesize (a) IID partition.}\\[1mm]
\includegraphics[width=\columnwidth]{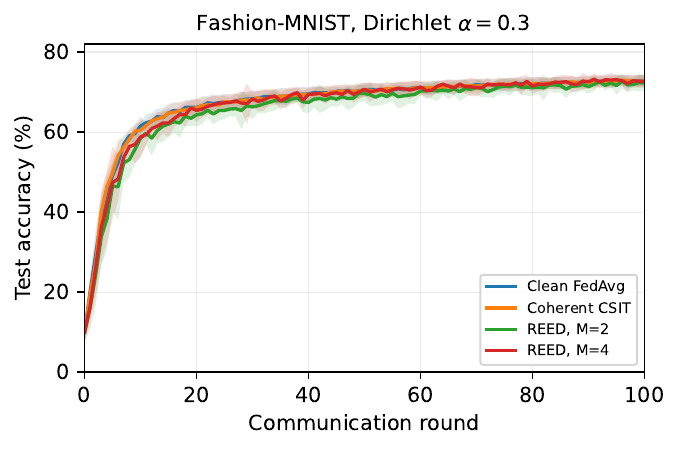}\\[-1mm]
{\footnotesize (b) Dirichlet \(\alpha=0.3\) partition.}
\caption{Chip-diverse REED on Fashion-MNIST at \(-10\)~dB effective receive SNR.}
\label{fig:chip_diversity}
\end{figure}

\begin{table*}[!t]
\centering
\caption{Round-100 test accuracy at \(-10\)~dB effective receive SNR. Values are mean \(\pm\) standard deviation over 10 independent runs. Gap columns report the REED gap relative to clean FedAvg in percentage points.}
\label{tab:round100_accuracy}
\renewcommand{\arraystretch}{1.08}
\setlength{\tabcolsep}{3.0pt}
\begin{tabular}{lcccccccc}
\toprule
Setting & Clean FedAvg & Coherent CSIT & REED \(M=1\) & REED \(M=2\) & REED \(M=4\) & Gap \(M=1\) & Gap \(M=2\) & Gap \(M=4\) \\
\midrule
MNIST, IID
& \(94.65\pm0.32\)
& \(94.71\pm0.32\)
& \(94.61\pm0.43\)
& --
& --
& \(-0.04\) pp
& --
& -- \\

MNIST, Dirichlet \(\alpha=0.3\)
& \(93.15\pm0.45\)
& \(93.16\pm0.48\)
& \(92.55\pm0.67\)
& --
& --
& \(-0.60\) pp
& --
& -- \\

Fashion-MNIST, IID
& \(75.04\pm0.94\)
& \(75.08\pm0.93\)
& \(74.78\pm0.95\)
& \(74.85\pm1.06\)
& \(75.10\pm1.05\)
& \(-0.26\) pp
& \(-0.18\) pp
& \(+0.06\) pp \\

Fashion-MNIST, Dirichlet \(\alpha=0.3\)
& \(72.83\pm1.52\)
& \(73.09\pm1.56\)
& \(69.66\pm3.55\)
& \(72.25\pm1.97\)
& \(72.61\pm1.50\)
& \(-3.17\) pp
& \(-0.58\) pp
& \(-0.21\) pp \\
\bottomrule
\end{tabular}
\end{table*}

\subsection{Baseline Comparison with \(M=1\)}

Figures~\ref{fig:single_shot_mnist} and~\ref{fig:single_shot_fashion} compare REED with \(M=1\) against clean FedAvg and coherent CSIT aggregation. On MNIST, REED closely follows both references under IID and heterogeneous partitions. At round 100, the gap relative to clean FedAvg is \(0.04\) percentage points in the IID setting and \(0.60\) percentage points under Dirichlet \(\alpha=0.3\), as reported in Table~\ref{tab:round100_accuracy}.

Fashion-MNIST provides a more sensitive test of noncoherent aggregation in this low-SNR regime. In the IID setting, REED with \(M=1\) remains close to clean FedAvg, with a round-100 gap of \(0.26\) percentage points. Under Dirichlet \(\alpha=0.3\), the \(M=1\) gap increases to \(3.17\) percentage points, creating a setting in which the variance-reduction effect of chip diversity can be observed. Therefore, we use Fashion-MNIST for the chip-diversity study below.

\subsection{Effect of Chip Diversity}

Figure~\ref{fig:chip_diversity} evaluates chip-diverse REED on Fashion-MNIST by increasing the number of independently faded paired observations per coordinate from \(M=1\) to \(M=2\) and \(M=4\). In the IID setting, the \(M=1\) curve is already close to clean FedAvg, so additional diversity produces only small changes. The round-100 gap changes from \(-0.26\) percentage points for \(M=1\) to \(-0.18\) for \(M=2\) and \(+0.06\) for \(M=4\), all within the observed run-to-run variability.

The heterogeneous Fashion-MNIST setting shows the benefit more clearly. At \(-10\)~dB, increasing the diversity from \(M=1\) to \(M=2\) improves the round-100 accuracy from \(69.66\%\) to \(72.25\%\), reducing the gap to clean FedAvg from \(3.17\) to \(0.58\) percentage points. With \(M=4\), the accuracy reaches \(72.61\%\), leaving a \(0.21\)-percentage-point gap. These results are consistent with the variance law in Section~\ref{sec:chip_reed}: independent energy observations reduce fading-induced self-noise, and the gain is most apparent when the low-SNR \(M=1\) perturbation is large enough to affect the learning dynamics.

Overall, REED supports FedAvg without instantaneous CSIT/CSIR or phase-coherent combining. REED with \(M=1\) is sufficient in the easier regimes considered here, while modest chip diversity substantially reduces the loss in the more sensitive heterogeneous Fashion-MNIST setting.

\FloatBarrier

\section{Conclusion}
\label{sec:conclusion}

This paper proposed REED, a simple paired-energy method for noncoherent signed aggregation in OTA-FL. By transmitting the positive and negative parts of each scalar update over paired orthogonal resource elements, REED forms signed aggregates from energy differences without instantaneous CSIT/CSIR, channel inversion, artificial offsets, or sign-only quantization.

We characterized the resulting estimator under Rayleigh fading and showed how chip diversity reduces fading-induced self-noise through independent energy observations. When incorporated into full-participation FedAvg, REED yields an unbiased aggregation perturbation and preserves the standard \(O(1/\sqrt{T})\) stationarity scaling under energy-scaled aggregation. Experiments at \(-10\)~dB effective receive SNR showed that REED with \(M=1\) closely tracks clean FedAvg in benign regimes, while modest chip diversity substantially reduces the gap in the more sensitive heterogeneous Fashion-MNIST setting.

REED offers a clean noncoherent alternative for signed model-update aggregation, trading additional resource elements or diversity branches for reduced channel-alignment requirements. Future directions include adaptive diversity allocation, decentralized REED-based consensus, and integration with compression or predictive update representations.

\appendices

\section{Proofs for the Scalar REED Estimator}
\label{app:reed_scalar_proofs}

Let
\[
S_+ \triangleq \sum_{k=1}^K \pos{u_k},
\qquad
S_- \triangleq \sum_{k=1}^K \negp{u_k}.
\]
We use one elementary fact repeatedly: if
$y\sim\mathcal{CN}(0,\nu)$, then
\[
\E |y|^2=\nu,
\qquad
\Var(|y|^2)=\nu^2.
\]
For a branch with chip weight $c\ge 0$, the matched-power normalization gives
\[
\sum_{k=1}^K h_{k,+}
\frac{1}{\mu_k}\sqrt{\eta c\,\pos{u_k}}e^{j\phi_{k,+}}
+z_+
\sim
\mathcal{CN}(0,\eta cS_+ +\sigma_z^2),
\]
and analogously the negative branch has variance
$\eta cS_-+\sigma_z^2$. This single Gaussian-energy identity proves all scalar
estimator statements below.

\subsection{Unbiasedness in Proposition~\ref{prop:reed_moments}}
Taking $c=1$ in the preceding identity,
\[
\E[|y_+|^2-|y_-|^2]
=(\eta S_+ +\sigma_z^2)-(\eta S_-+\sigma_z^2)
=\eta(S_+-S_-).
\]
Therefore
\[
\E[\hat s]=S_+-S_-=
\sum_{k=1}^K(\pos{u_k}-\negp{u_k})=
\sum_{k=1}^K u_k .
\]

\subsection{Variance in Proposition~\ref{prop:reed_moments}}
The two resource elements are independent, so
\[
\Var(\hat s)=
\frac{(\eta S_+ +\sigma_z^2)^2+(
\eta S_-+\sigma_z^2)^2}{\eta^2}.
\]
Expanding gives
\[
\Var(\hat s)=S_+^2+S_-^2+
\frac{2\sigma_z^2}{\eta}(S_++S_-)+
\frac{2\sigma_z^4}{\eta^2}.
\]
Since $S_++S_- =\sum_{k=1}^K |u_k|$, this is
\eqref{eq:reed_variance}.

\subsection{Proof of Proposition~\ref{prop:chip_reed_moments}}
For chip $m$, the preceding calculation with $c=c_m$ yields
\[
\E[|y_{m,+}|^2-|y_{m,-}|^2]=\eta c_m(S_+-S_-),
\]
so normalization by $\eta C_M$ gives
$\E[\hat s_M]=S_+-S_-$. Independence across chips and branches gives
\[
\Var(\hat s_M)=
\frac{1}{\eta^2C_M^2}
\sum_{m=1}^M
\left[(\eta c_mS_+ +\sigma_z^2)^2
+(\eta c_mS_-+\sigma_z^2)^2\right].
\]
Expanding the sum and using $C_M=\sum_m c_m$ gives
\[
\Var(\hat s_M)
=
\frac{\sum_{m=1}^M c_m^2}{C_M^2}(S_+^2+S_-^2)
+
\frac{2\sigma_z^2}{\eta C_M}(S_++S_-)
+
\frac{2M\sigma_z^4}{\eta^2C_M^2},
\]
which is \eqref{eq:chip_reed_variance} because
$S_++S_- = \sum_{k=1}^K |u_k|$.

\section{Convergence Proofs}
\label{app:convergence_proof}

The argument follows the standard smooth nonconvex FedAvg/OTA-FL descent template; see,
e.g., \cite{Zhu2024OTAFLOpt,Cao2022JSAC_AirFedAvgPowerControl,Li2020FedAvgNonIID,Li2020FedProx}.
We therefore make explicit only the REED-specific ingredients:
(i) REED induces an additive \emph{unbiased} aggregation error with an explicit
second-moment bound, and
(ii) under a per-client energy budget, the common gain can be scheduled so that
the REED distortion scales as \(O(\beta^2)\).

\subsection{Proof of Lemma~\ref{lem:reed_vector}}

Apply REED coordinate-wise with equal weights
\[
u_{k,j}^t=\frac{1}{K}[\Delta_k^t]_j.
\]
For each coordinate, Proposition~\ref{prop:reed_moments} or Proposition~\ref{prop:chip_reed_moments} gives an unbiased scalar estimate of \([\bar\Delta^t]_j\). Hence
\[
\E\!\left(\varepsilon^t \,\middle|\, \cF_t,\{\Delta_k^t\}_{k=1}^K\right)=0,
\qquad
\E[\varepsilon^t\mid \cF_t]=0.
\]

We next bound the accumulated distortion across coordinates. For \(M=1\), Proposition~\ref{prop:reed_moments} gives
\[
\E[(\varepsilon_j^t)^2]
=
(S_{+,j}^t)^2+(S_{-,j}^t)^2
+\frac{2\sigma_z^2}{\eta}\sum_{k=1}^K |u_{k,j}^t|
+\frac{2\sigma_z^4}{\eta^2},
\]
where
\[
S_{+,j}^t=\sum_{k=1}^K [u_{k,j}^t]_+,
\qquad
S_{-,j}^t=\sum_{k=1}^K [u_{k,j}^t]_- .
\]
The self-noise term satisfies
\[
\begin{aligned}
\sum_{j=1}^d \big((S_{+,j}^t)^2+(S_{-,j}^t)^2\big)
&\le
\sum_{j=1}^d \Big(\sum_{k=1}^K |u_{k,j}^t|\Big)^2 \\
&= \Big\|\sum_{k=1}^K |u_k^t|\Big\|_2^2
\le \Big(\sum_{k=1}^K \|u_k^t\|_2\Big)^2  \\
&=
\Big(\frac{1}{K}\sum_{k=1}^K \|\Delta_k^t\|_2\Big)^2
\le (\beta QG)^2,
\end{aligned}
\]
where the last step uses Assumption~(A3). The signal--noise term is bounded by
\[
\sum_{j=1}^d\sum_{k=1}^K |u_{k,j}^t|
=
\frac{1}{K}\sum_{k=1}^K \|\Delta_k^t\|_1
\le
\frac{\sqrt d}{K}\sum_{k=1}^K \|\Delta_k^t\|_2
\le \sqrt d\,\beta QG .
\]
Combining these bounds over all coordinates gives \eqref{eq:sigma_air}.

For chip-diverse REED, Proposition~\ref{prop:chip_reed_moments} multiplies the self-noise contribution by \(\sum_m c_m^2/C_M^2\), replaces the signal--noise coefficient by \(2\sigma_z^2/(\eta C_M)\), and replaces the pure noise contribution by \(2M\sigma_z^4/(\eta^2 C_M^2)\) per coordinate. Applying the same two vector bounds above gives \eqref{eq:chip_sigma_air}. This proves Lemma~\ref{lem:reed_vector}.

\subsection{Proof of Lemma~\ref{lem:eta_beta_scaling}}

With average channel-power normalization, the total energy used by client \(k\) on coordinate \(j\), summed over all chip pairs, is
\[
\sum_{m=1}^M\big(|a_{k,j,m,+}^t|^2+|a_{k,j,m,-}^t|^2\big)
=
\frac{\eta C_M}{K\mu_k^2}\,|[\Delta_k^t]_j|.
\]
Therefore
\[
\frac{1}{d}\sum_{j=1}^d\sum_{m=1}^M
\big(|a_{k,j,m,+}^t|^2+|a_{k,j,m,-}^t|^2\big)
=
\frac{\eta C_M}{K\mu_k^2 d}\,\|\Delta_k^t\|_1.
\]
Assumption~(A3) gives
\[
\|\Delta_k^t\|_1
\le
\sqrt d\,\|\Delta_k^t\|_2
\le
\sqrt d\,\beta QG.
\]
Thus, the average energy constraint is satisfied whenever
\[
\eta
\le
\frac{E_k K\sqrt d\,\mu_k^2}{C_M\beta QG}
\qquad\text{for every }k,
\]
which yields \eqref{eq:eta_choice}. For fixed system parameters and fixed chip weights, \(\eta=\Theta(1/\beta)\). Substituting this scaling into \eqref{eq:sigma_air} or \eqref{eq:chip_sigma_air} gives \(\sigma_{\mathrm{air}}^2=O(\beta^2)\), since the three terms scale as \(O(\beta^2)\), \(O(\beta/\eta)\), and \(O(1/\eta^2)\), respectively. This proves Lemma~\ref{lem:eta_beta_scaling}.

\subsection{Proof of Theorem~\ref{thm:conv}}
\label{app:proof_thm_conv}

The server update consists of the ideal FedAvg increment plus the REED aggregation
error. Thus,
\[
w^{t+1}=w^t+\widehat{\Delta}^t,
\qquad
\widehat{\Delta}^t=\bar{\Delta}^t+\varepsilon^t,
\qquad
\bar{\Delta}^t=\frac{1}{K}\sum_{k=1}^K \Delta_k^t.
\]
By Lemma~\ref{lem:reed_vector},
\[
\E[\varepsilon^t\mid \cF_t]=0,
\qquad
\E\|\varepsilon^t\|^2\le \sigma_{\mathrm{air}}^2.
\]

\paragraph{Step 1: one-step smoothness bound.}
By \(L\)-smoothness,
\begin{equation}
F(w^{t+1})
\le
F(w^t)
+\langle \nabla F(w^t),\widehat{\Delta}^t\rangle
+\frac{L}{2}\|\widehat{\Delta}^t\|^2.
\label{eq:app_smooth_step}
\end{equation}
Taking expectation and using
\[
\E\langle \nabla F(w^t),\varepsilon^t\rangle=0,
\qquad
\E\langle \bar{\Delta}^t,\varepsilon^t\rangle=0,
\]
gives
\begin{equation}
\E[F(w^{t+1})]
\le
\E[F(w^t)]
+\E\langle \nabla F(w^t),\bar{\Delta}^t\rangle
+\frac{L}{2}\E\|\bar{\Delta}^t\|^2
+\frac{L}{2}\sigma_{\mathrm{air}}^2.
\label{eq:app_descent_split}
\end{equation}

\paragraph{Step 2: control the drift term.}
Define
\[
\bar g_q^t \triangleq \frac{1}{K}\sum_{k=1}^K g_{k,q}^t,
\qquad
\bar\nabla_q^t \triangleq \frac{1}{K}\sum_{k=1}^K \nabla f_k(w_{k,q}^t),
\]
so that
\[
\bar\Delta^t = -\beta\sum_{q=0}^{Q-1} \bar g_q^t.
\]
By Assumption (A2),
\[
\E[\bar g_q^t \mid \{w_{k,q}^t\}_k]=\bar\nabla_q^t.
\]
Define the local-drift term
\[
\delta_q^t \triangleq \bar\nabla_q^t-\nabla F(w^t)
=
\frac{1}{K}\sum_{k=1}^K\bigl(\nabla f_k(w_{k,q}^t)-\nabla f_k(w^t)\bigr).
\]
Then
\[
\E\langle \nabla F(w^t),\bar\Delta^t\rangle
=
-\beta\sum_{q=0}^{Q-1}
\E\Bigl[\|\nabla F(w^t)\|^2+\langle \nabla F(w^t),\delta_q^t\rangle\Bigr].
\]
Using \(\langle a,b\rangle\ge -\tfrac12\|a\|^2-\tfrac12\|b\|^2\),
\begin{equation}
\E\langle \nabla F(w^t),\bar\Delta^t\rangle
\le
-\frac{\beta Q}{2}\E\|\nabla F(w^t)\|^2
+\frac{\beta}{2}\sum_{q=0}^{Q-1}\E\|\delta_q^t\|^2.
\label{eq:app_innerprod_bound}
\end{equation}

By \(L\)-smoothness and Assumption (A3),
\[
\|w_{k,q}^t-w^t\|
=
\Big\|\beta\sum_{r=0}^{q-1} g_{k,r}^t\Big\|
\le
\beta q G,
\]
hence
\[
\|\nabla f_k(w_{k,q}^t)-\nabla f_k(w^t)\|
\le
L\beta q G.
\]
Therefore \(\|\delta_q^t\|\le L\beta qG\), and
\begin{equation}
\sum_{q=0}^{Q-1}\E\|\delta_q^t\|^2
\le
L^2\beta^2 G^2 \sum_{q=0}^{Q-1} q^2
\le
L^2\beta^2 Q^3 G^2.
\label{eq:app_drift_sum}
\end{equation}
Combining \eqref{eq:app_innerprod_bound} and \eqref{eq:app_drift_sum} gives
\begin{equation}
\E\langle \nabla F(w^t),\bar\Delta^t\rangle
\le
-\frac{\beta Q}{2}\E\|\nabla F(w^t)\|^2
+\frac{L^2}{2}\beta^3 Q^3 G^2.
\label{eq:app_innerprod_final}
\end{equation}

\paragraph{Step 3: bound \(\E\|\bar\Delta^t\|^2\).}
Let
\[
\bar\xi_q^t \triangleq \bar g_q^t-\bar\nabla_q^t.
\]
Using \(\|\sum_q x_q\|^2\le Q\sum_q \|x_q\|^2\),
\[
\E\|\bar\Delta^t\|^2
=
\beta^2 \E\Big\|\sum_{q=0}^{Q-1}\bar g_q^t\Big\|^2
\le
\beta^2 Q \sum_{q=0}^{Q-1}\E\|\bar g_q^t\|^2.
\]
Next,
\[
\|\bar g_q^t\|^2
\le
2\|\bar\nabla_q^t\|^2+2\|\bar\xi_q^t\|^2
\le
4\|\nabla F(w^t)\|^2+4\|\delta_q^t\|^2+2\|\bar\xi_q^t\|^2.
\]
Hence
\begin{equation}
\E\|\bar\Delta^t\|^2
\le
\beta^2 Q \sum_{q=0}^{Q-1}
\E\Bigl[4\|\nabla F(w^t)\|^2+4\|\delta_q^t\|^2+2\|\bar\xi_q^t\|^2\Bigr].
\label{eq:app_delta2_expand}
\end{equation}

Under full participation and Assumption (A4), the averaged gradient noise satisfies
\begin{equation}
\E\|\bar\xi_q^t\|^2 \le \frac{\sigma_g^2}{K},
\label{eq:app_noise_avg}
\end{equation}
provided client gradient noises are conditionally independent or uncorrelated given the iterates.
Using \eqref{eq:app_drift_sum} and \eqref{eq:app_noise_avg} in \eqref{eq:app_delta2_expand} yields
\begin{equation}
\E\|\bar\Delta^t\|^2
\le
4\beta^2 Q^2 \E\|\nabla F(w^t)\|^2
+4L^2\beta^4 Q^4 G^2
+\frac{2\beta^2 Q^2}{K}\sigma_g^2.
\label{eq:app_delta2_final}
\end{equation}

\paragraph{Step 4: combine and telescope.}
Substitute \eqref{eq:app_innerprod_final} and \eqref{eq:app_delta2_final} into \eqref{eq:app_descent_split}:

\begin{equation}
\begin{aligned}
\E[F(w^{t+1})]
\le\;&
\E[F(w^t)]
-\frac{\beta Q}{2}\E\|\nabla F(w^t)\|^2
+\frac{L^2}{2}\beta^3 Q^3 G^2  \\
&+\frac{L}{2}\Big(
4\beta^2 Q^2 \E\|\nabla F(w^t)\|^2
+4L^2\beta^4 Q^4 G^2 \\
&\hspace{2.5cm}
+\frac{2\beta^2 Q^2}{K}\sigma_g^2
\Big)
+\frac{L}{2}\sigma_{\mathrm{air}}^2 .
\end{aligned}
\label{eq:app_descent_combined}
\end{equation}
If \(\beta\le 1/(8LQ)\), then \(2L\beta^2Q^2\le \beta Q/4\), so
\[
-\frac{\beta Q}{2}+2L\beta^2Q^2 \le -\frac{\beta Q}{4}.
\]
Therefore,

\begin{equation}
\begin{aligned}
\frac{\beta Q}{4}\E\|\nabla F(w^t)\|^2
\le\;&
\E[F(w^t)-F(w^{t+1})]
+\frac{L^2}{2}\beta^3 Q^3 G^2  \\
&+2L^3\beta^4 Q^4 G^2
+\frac{L\beta^2 Q^2}{K}\sigma_g^2
+\frac{L}{2}\sigma_{\mathrm{air}}^2 .
\end{aligned}
\label{eq:app_key_ineq}
\end{equation}
The remaining argument is a standard telescoping step. Summing
\eqref{eq:app_key_ineq} over \(t=0,\dots,T-1\) and using the lower bound
\(F(w^T)\ge F_\star\), we obtain

\begin{equation}
\begin{aligned}
\frac{\beta Q}{4}
\sum_{t=0}^{T-1}\E\|\nabla F(w^t)\|^2
\le\;&
F(w^0)-F_\star  \\
&+T\Big(
\frac{L^2}{2}\beta^3 Q^3 G^2
+2L^3\beta^4 Q^4 G^2  \\
&\hspace{1.7cm}
+\frac{L\beta^2 Q^2}{K}\sigma_g^2
+\frac{L}{2}\sigma_{\mathrm{air}}^2
\Big).
\end{aligned}
\end{equation}
Dividing by \((\beta Q/4)T\) yields \eqref{eq:conv_bound}. \qed
\bibliographystyle{IEEEtran}
\bibliography{IEEEabrv,references}

\end{document}